\documentclass[prl,twocolumn]{revtex4}
\usepackage{graphicx}
\usepackage{bm}

\begin{document}

\title{Non Classical Rotational Inertia Fraction in a One Dimensional Model of Supersolid}

\author{
N\'estor Sep\'ulveda$^1$, Christophe Josserand$^{2,3}$,  and Sergio Rica$^{1,4}$}
\affiliation{$^1$ Laboratoire de Physique Statistique de l'Ecole
normale sup\'erieure, 24
Rue Lhomond, 75231 Paris Cedex 05, France.\\
$^2$Institut Jean Le Rond D'Alembert,
 UMR 7190 CNRS-UPMC, 4 place Jussieu, 75005 Paris, France.\\
$^3$Kavli Institute for Theoretical Physics, University of California, Santa Barbara, CA 93106,USA\\
$^4$Departamento de F\'\i sica, Universidad de Chile, Blanco Encalada 2008, Santiago, Chile.}

\begin{abstract}
We study the rotational inertia of  a model of supersolid in the frame of the mean field Gross-Pitaevskii theory in one space dimension. We discuss the ground state of the model and  the existence of a non classical inertia (NCRI) under rotation that models an annular geometry. An explicit formula for the 
NCRI is deduced. It depends on the density profil of the ground state, in full agreement with former 
theories. We compare the 
NCRI computed through this theory with direct numerical simulations of rotating 1D systems.

\end{abstract}

\maketitle

\section{Introduction}

Since pioneering works by Andreev and Lifshitz\cite{andreev}, Chester\cite{chester}, Leggett\cite{leggett} and others, supersolids have been dreamt up as a kind of Bose-Einstein condensation of defects, vacancies or interstitials. They would achieve a coherent state that could allow a matter flow trough the crystal. Although the quest for a supersolid state over the last 40 years has failed\cite{meisel}, the context has totally changed with the recent experiments by Kim and Chan\cite{chan04a,chan04b,chan06}. In these experiments, solid Helium$^4$ fills a torsional oscillator under small oscillations and the rotational frequency is measured. Surprisingly, the rotational inertia shows a drop below few tenths of Kelvin. This non-classical rotational inertia (NCRI) is believed to be the signature of the transition of a fraction of the solid into a supersolid state. The situation has become puzzling as other experiments have been performed. Thus, although the drop of the moment of inertia has been confirmed, crystal annealing was shown to lower dramatically the amplitude of this NCRI \cite{RiRe06,RiRe07}. Similarly, solid Helium submitted to pressure gradient could not flow except when large grain boundaries where present in the sample \cite{Grey77,Bonfait89,Bali06}. Moreover, the responses of solid $^4$He to a localized pressure jump
 presented no evidence of superflow in the solid\cite{Beamish05,Beamish06} .The experimental context 
 presents thus apparent contradictions between NCRI measurements and pressure driven flows with the role of disorder (through vacancies, grain boundaries...) to be elucidated.
On the other hand the theoretical framework for describing supersolids presents also fundamental puzzles (see the recent review of Prokof'ev\cite{Prok07} where the influence of the disorder is 
particularly discussed). Beside Penrose and Onsager argument\cite{po}, Monte-Carlo models claim that a perfect crystal cannot exhibit supersolid behavior\cite{CeBu04,PrSv05}. However, the account for exchange processes between neighboring atoms \cite{leggett,Leg04}, the densities and the role of the vacancies in the dynamics raise still fundamental questions on the
existence and the nature of the supersolidity (see \cite{And05,Rea05} for instance).

An alternative issue consists of using the Gross-Pitaevskii (GP) model\cite{pit} to describe the dynamics 
of a quantum solid, as proposed in 1994 by Pomeau and Rica\cite{pomric}. 
The original GP equation\cite{pit} is used, with a roton minimun in the dispersion 
relation, where the ground state  exhibits a first order phase transition to a crystalline state. 
However, the assumptions underlying the GP equation are not strictly speaking valid for Helium although this equation is believed to give a good qualitative description of superfluid Helium. Therefore this model, even crude in its basic structure, is at least a good testbed for theories of supersolids, that are still in a state of uncertainty. 
In Ref. \cite{ss1a,ss1b}, two of us have developped the theory for the long wave perturbations of this model of supersolid and we have shown that this model was able to conciliate the apparent experimental contradiction discussed above.
In the present paper, we study the one-dimensional (1D) version of this model. Beside the simplicity of the 1D approach, which then allows precise determination of the
crystal structure, the 1D limit is particularly interesting since it can model to some extent an annular geometry.

\section{The model}

The starting point is the original GP equation\cite{pit} for the complex wavefunction $\psi(x,t)$ in one space dimension: 
\begin{equation}
    i\hbar \frac{\partial \psi}{\partial t} = 
    -\frac{\hbar^{2}}{2m} \frac{ \partial^2 \psi}{\partial x^2}  + \psi\int_{-\infty}^{\infty} U(|x-y|) |\psi(y)|^2 {\mathrm{d}}y 
    \mathrm{,}
\label{nls.org}
\end{equation}
where $U(s)$ is the two body potential depending on the relative distance. 
The potential $U(s)$ should satisfy $$0<\int _{-\infty}^\infty U(s) ds <\infty,$$ for stability and its
Fourier transform: 
\begin{equation}
\hat U_k= \int^\infty_{-\infty} U(s) e^{i k s} ds \label{fourier}
\end{equation}
has to be bounded for all $k$. Moreover, as we will see later, we shall require also that  the Fourier transform $\hat U_k$
becomes negative at some $k_c$ to allow roton crystallization. 

This dynamics conserves the Hamiltonian (or the energy, following $\partial_t \psi= \frac{\delta H}{\delta \psi^*}$)
$$H =   \frac{\hbar^{2}}{2m} \int_{-\infty}^\infty  \left| \partial_x \psi \right|^2   {\mathrm{d}}x \, +\, \frac{1}{2}\int^\infty_{-\infty} \int^\infty_{-\infty} U(|x-y|) |\psi(y)|^2 |\psi(x)|^2 {\mathrm{d}}y  {\mathrm{d}}x   \mathrm{,}$$
 the number of particles 
 $$N =  \int^\infty_{-\infty}  |\psi(x)|^2  {\mathrm{d}}x $$
  and the linear momentum 
  $$ P = -\frac{i \hbar}{2} \int_{-\infty}^\infty  (\psi^*\partial_x\psi - \psi \partial_x\psi ^*)  {\mathrm{d}}x.$$ 
According to the energy, the ground-state solution is real since any non uniform phase increases its energy. 

The dynamics exhibits indeed an homogenous and 
stationnary solution $\psi_0=\sqrt{n_0}  e^{-i\frac{E_0}{\hbar} t} $, with $n_0$ the mean 1D density and $E_0 =  n_0 \int _{-\infty}^\infty U(s) ds $. This solution is 
stable and can also be the ground state for small enough $n_0$ as suggested by the  Bogoliubov spectrum of the perturbations\cite{bog} (see below):  
$$\hbar \omega_k = \sqrt{ \left( \hbar^2 k^2/2 m \right)^2  +    \left( \hbar^2 k^2/m\right) n\hat U_k}$$

Assuming that the potential scales like $U_0$ and possesses a single length scale $a$, the spectrum depends then only on a single dimensionless parameter\cite{pomric}: 
$$\Lambda=  n_0 \frac{m   a^2}{\hbar^2}   \hat U_0,$$ 
with $ \hat U_0  = \int _{-\infty}^\infty U(s) ds \sim U_0 a $.
For some analytical results and for the numerics later on, we choose the soft core interaction, with no loss of generalities:
\begin{equation}
U(|x-y|) = U_0 \theta(a-|x-y|)\label{U0},\end{equation} with $\theta(\cdot)$ the Heaviside function. 
The Fourier transform of this special interaction potential is 
$$\hat U_k = 2 U_0 \frac{ \sin( k a)}{k }=\hat U_0 \frac{ \sin( k a)}{k }.$$ 

It should also be noticed that the special choice of the potential (\ref{U0}) is purely of practical interest because it is easy to implement in some numerical schemes and can be easily used for variational estimates. Other functions whose fourier transform would be negative for a wavenumber 
domain (strictly bigger than zero), would show similar properties. Among them are the classical two 
bodies atomic potential with strong repulsion at short scale and a slight attraction for large scale or a potential $\hat U_k$ 
choosen in such a way that the Bogoliubov dispersion relation matches the Landau spectrum with the right values of the speed of sound $c$, and the three roton parameters\cite{landau}.

With $x'=x/a$, 
$t'= \frac{ma^2}{\hbar}t$ and $\psi'= \sqrt{n_0} \psi$, the dimensionless GP equation for the Heaviside interaction (we drop the primes hereafter) reads:
\begin{equation}
    i \frac{\partial \psi}{\partial t}= 
    -\frac{1}{2} \frac{ \partial^2 \psi}{\partial x^2}  + \frac{\Lambda}{2}   \psi(x,t)\int_{x-1}^{x+1} |\psi(y)|^2 {\mathrm{d}}y 
    \mathrm{.}
\label{nls1}
\end{equation}

Finally, we emphasize that an annular geometry can be simplified into a 1D system by considering periodic boundary condition $\psi(x,t) = \psi(x+L,t)$ ($L$ is dimensionless) and by assuming that the transverse structure of the solid can be neglected.
We then define the energy and number of particles densities:
\begin{eqnarray}
   {\mathcal E }&= & \frac{1}{2L} \int_{0}^{L} \left(
     \left| \psi_x\right|^2   +    \frac{\Lambda}{2} |\psi|^2 \int^{x+1}_{x-1}  |\psi(y)|^2 {\mathrm{d}}y\right)  {\mathrm{d}}x \label{densenergy}
    \mathrm{,}
\label{energy}\\
\rho&=&n_0=  \frac{1}{L}\int_{0}^{L}  |\psi(x,t)|^2 {\mathrm{d}}x . \label{density}
\end{eqnarray}

\section{Ground state}

As shown in \cite{pomric}, for low $\Lambda$ the ground 
state is a superfluid (without positional order). However above a critical value, $\Lambda_c $ the ground state shows a periodic modulation of the density in space. Although in two and three space dimensions the transition is first order as $\Lambda$ 
increases, it  is supercritical (second order) in one space dimension\footnote{With the
potential constructed in (\ref{U0}) the critical value is $\Lambda_c = 21.05 \dots.$}.
The periodic structure arising from the instability can be analytically estimated through a variational 
approach for a fixed wavelength $\lambda$ at least in two regimes: close to the transition and for large 
$\Lambda$.

If $\Lambda \gtrsim \Lambda_c$, a weak amplitude developement of wanumber $k$ and 
normalized to unity reads:
\begin{equation}
\psi (x) = \frac{1}{\sqrt{1+ 2 |A|^2}}\left(1 + A e^{i k x}  + A^*e^{-i k x} \right).
\label{trial1}
\end{equation}
Minimizing the energy of such solution gives:

\begin{equation}|A|^2 =  - \frac{k^2 + 4 \Lambda \hat U_{k}/\hat U_{0}}{2 ( k^2 + \Lambda(\hat U_{2k}-4 \hat U_{k})/\hat U_{0})} 
\label{weakamp}\end{equation}

and the following wave-number selection $k_c a = 4.078\dots$. The amplitude for this wave number $k_c$ follows
$$ |A_c|^2 =(\Lambda-\Lambda_c) \frac{-4{\rm sin}(k_c)}{4k_c(k_c^2+\Lambda_c {\rm cos}^2(k_c)/3}). $$
while for $k\sim k_c$, $|A_c|^2-|A|^2 \propto (k-k_c)^2$.

In the large $\Lambda$ limit, the density exhibits strongly nonlinear strucures since the potential energy in (\ref{densenergy}) requires small $\psi$ while the mass 
normalization (\ref{density}) forbids $\psi$ to be small every-where. Therefore the energy 
minimization leads to a periodic structure with zones where $\psi \approx 0$ balancing zones where 
$\psi \gg 1.$ 
In the $\Lambda \rightarrow \infty$ limit, Ref. \cite{mamandine} showed that $\psi \neq 0 $ only in a small zone $x\in (-\delta,\delta)$ 
of the whole period $(0,\lambda)$.  The Euler-Lagrange condition deduced from (\ref{densenergy}) together with  (\ref{density}) leads to the Hemholtz equation in the domain $(-\delta,\delta)$ : $-\psi''(x) =  \mu \psi$. Finally the minimization of the energy gives $\delta$ and the wave number $\lambda$ of the periodic structure. 
Following this approach, we sketch now an estimate of the ground state for finite $\Lambda \gg 1$.
%We consider the energy (\ref{densenergy}) and normalization (\ref{density}) in a single period.
We use the trial function for a single period: 
\begin{equation}
\psi(x) = \sqrt{\frac{\lambda}{\delta}} \cos\left(\frac{\pi x}{2\delta}\right)
\label{trial2}
\end{equation}
in $x \in[-\delta,\delta]$ and zero elsewhere, that satisfies exactly the normalization condition (\ref{density}). Introducing this trial function into the  energy (\ref{densenergy}), we obtain: $   {\mathcal E } =    {\mathcal E }_1 +    {\mathcal E }_2 +    {\mathcal E }_3,$ 
with the kinetic energy 
$$   {\mathcal E }_1 = \frac{\pi^2}{8\delta^2}, $$
the self interaction of a pulse with itself 
$$   {\mathcal E }_2 =  \frac{\Lambda\lambda}{4}, $$
and the nontrivial interaction of a pulse with its two near neighbors
$${\mathcal E}_3=   \frac{\Lambda}{2} \int_{\lambda-1-\delta}^\delta \psi(x)^2 \int_{-\delta}^{x+1-\lambda}  \psi(y)^2 dy \, dx .$$
This energy ${\mathcal E}_3$ is not zero only if $\lambda < 1 + 2 \delta$ (and naturally, we have $\lambda> 2 \delta$).

\begin{figure}[hc]
\begin{center}
\centerline{\includegraphics[width=8cm]{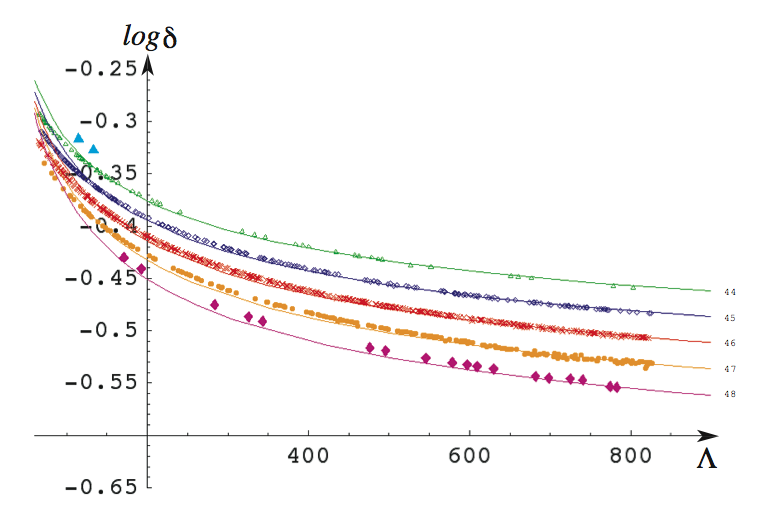} }
\caption{ \label{deltavsLambda} Plot of  $log_{10}\delta $ as a function of $\Lambda$ for different 
$\lambda$. The curves correspond to formula (\ref{Lambda1}) while the points come from numerical simulations. The total size of the system is $64$ units and the range of the interaction is  $a=16$. The system displays a number of cells varying from 44 to 48 (identified at the right hand side of the figure). The respectives $\lambda$ vary thus from $4/3$ to $16/11$.}
\end{center}
\end{figure}

The energy ${\mathcal E}$ may be understood as a function of $\delta$ and of the periodicity length $\lambda$. Then, 
the variation of total energy in the $(\delta,\lambda)$ plane shows the existence of a global 
minimum and a saddle for large enough $\Lambda$. As $\Lambda$ decreases, the saddle and
 the global minimum collide, and we obtain a pure monotonic energy landscape in the 
 $(\delta,\lambda)$ plane, leading to both $\lambda$ and $\delta$ to infinity as minimizer. On the other hand, when 
 $\Lambda\rightarrow \infty$ the global minimum moves to
 $(\delta,\lambda) \rightarrow (0,1)$.
This selection mechanism holds for an infinite domain where the wavelength $\lambda$ can vary 
continuously. In a finite domain, $\lambda $ can only take discrete values related to the 
number $N_c$ of unitary cells, $\lambda = L/ N_c$. Numerical simulations suggest also the existence 
of a large energy barrier between minimizers with different number of cells $N_c$ for large
$\Lambda$. Thus, for a given domain size $L$, the energy, as function of $(\delta,\lambda)$, 
is now described by a discrete set of energy functions of $\delta$ for each available $\lambda$ satisfying $\lambda = L/ N_c$. One has now to minimize each energy with respect to $\delta$.
For small $\Lambda$ (typically smaller than $\Lambda_c$)
none of these functions have minima. For large $\Lambda$ on the other hand, there is a finite band 
of $\lambda$ for which the energy admitts a minimum as $\delta$ varies.
The minimization of this energy respect to $\delta$ provides relations among  $\delta$ and $\Lambda$ with the wavelength $\lambda$ as a fixed parameter. To determine the global minimum and to avoid
further algebraic difficulties, we introduce the new variable: $z = \pi  (\lambda -1)/\delta$
where $0\leq z\leq 2\pi$ for our problem (in particular, $z>2\pi$ means that the peaks do not interact
one with another). Minimizing the energy gives a relation for $\Lambda=\Lambda(\lambda,z)$: 
\begin{equation}
\Lambda=\frac{4 \pi ^2 z}{ \lambda  (\lambda -1) ^2  ((2 \pi -z)
   (\cos (z)+2)+3 \sin (z))}.\label{Lambda1}
\end{equation} 
Fig. \ref{deltavsLambda} shows $\delta$ versus $\Lambda$ for different values of $\lambda$.
The analytical curves are shown together with the results of direct numerical simulations described below.

Finally,  an exponential ``boundary layer" correction developps near $x=\pm \delta$ where the nonlinear
term in eq. (\ref{nls1}) cannot be neglected, as noticed by \cite{mamandine}.
In the limit of large $\Lambda$, where similarly $\lambda \rightarrow 1$, the nonlinear term of
eq. (\ref{nls1}) gives near $x=\delta$:
  
$$\lim_{\lambda\rightarrow 1} \int_{x-1}^{x+1} \psi(y)^2 dx = Cte+\lim_{\lambda\rightarrow 1} \int_{\lambda-\delta}^{x+1} \psi(y)^2 dx \approx  Cte+ \frac{{\pi }^2 }{12\,{\delta }^3} {\left( x - \delta  \right) }^3+ {\mathcal O}(( x - \delta   )^4)$$

The ground state is thus modified into $\psi(x) +\varphi(x)$, where $\psi(x)$ is the trial function 
(\ref{trial2}), and $\varphi $ satisfies a linear Schr\"odinger equation: 
$$ -\frac{1}{2}\varphi''(x) + \frac{\Lambda}{2} V(x)  \varphi(x)=0,$$ 
where  the first nontrivial term for the potential reads $V(x) =  \frac{{\pi }^2\,\lambda}{12\,{\delta }^3} {\left( x - \delta  \right) }^3$. The solution may be  computed directly in term of Bessel functions : $\varphi(x) = K_* \sqrt{ x} K_{1/5}\left(\frac{\pi}{5}  \sqrt{\frac{ \Lambda \lambda }{3 \delta^3}} (x-\delta) ^{5/2} \right).$ where the constant $K_* $ results from the matching of the  exponentially small boundary layer and the trial function. One can also expand this solution  {\it via} a WKB approxomation
$\varphi(x) = K_* e^{-\sqrt{\Lambda} S(x)}$ with $S(x) = S_0(x) + \frac{1}{\sqrt{\Lambda} } S_1(x)+\dots $.
We obtain then  $S_0(x) = \int \sqrt{V(x)} dx$  and $S_1(x) = \frac{1}{2} \log(S_0'(x)) $ and therefore $$\varphi(x) = \frac{K_*}{\sqrt{S_0'(x)}} e^{-\sqrt{\Lambda} S_0(x)}$$ with  $S_0(x) =  \frac{{\pi }\,\sqrt{\lambda/3}}{5\,{\delta }^{3/2}} {\left( x - \delta  \right) }^{5/2}$.

\section{Non-classical moment of inertia in superfluids and supersolids.}

The precise estimation of the ground state is in fact crucial to describe the supersolid features of the model.
Indeed,  we have obtained in Refs. \cite{ss1a,ss1b}, using the homogenization technique \cite{lions}, an expression for the effective or superfluid density matrix $\varrho^{ss}_{ik}$ deduced from the density profile of the crystal. 

We shall in fact explore the low excited states around the ground state, described by the knowledge of the 
crystal density $\rho_0(x)$. The change of  energy for phase variations gives:
\begin{equation}
 \Delta E  =  \frac{1}{2} \int  \rho_0(x)  \left( \frac{ \partial \phi}{\partial x}  \right)^2 dx
    \mathrm{.}
\label{kin}
\end{equation} 

and $\phi$ is determined by minimizing $\Delta E$, that correspo, the Euler-Lagrange condition: 
\begin{equation}
 \frac{\partial }{\partial x}  \left( \rho_0(x) \frac{\partial \phi}{\partial x} \right) = 0.
 \label{cont}
\end{equation} 
under the appropriate boundary conditions.
As shown by Leggett \cite{leggett}, for a periodic $\rho_0(x)$ under  rotation, $\Delta E  $ is lower than 
that of a rigid solid rotation, which indicates that superfluidity is present. 

In Refs. \cite{ss1a, ss1b} we have obtained an expression for the energy variation in three space dimensions $ \Delta E  =  \frac{\hbar^{2}}{2m} \int  {\varrho}^{ss}_{ik} \partial_i\phi \partial _k \phi    d^3x$, where ${\varrho}^{ss}_{ik}$ is the effective or 
superfluid density matrix. It can be explicitely expressed using a solution of a partial differential equation in the unit cell $V$
of the solid, following:

\begin{eqnarray}
 {\varrho}^{ss}_{ik} &= & n  {\delta}_{ik}  -   {\varrho}_{ik} \nonumber\\
 {\varrho}_{ik} &=& \frac{1}{V} \int_{V}  \rho_{0}({\bm r})\,  {\bm \nabla} {K}_{i}\cdot   {\bm \nabla} K_{k}   \, d{\bm r}. \label{varrhomatrix}
 \end{eqnarray}
 The vector $K_i$ is a periodic function in the unit cell $V$ that is solution of $\nabla_i\rho_{0} + 
    {\bm{\nabla}}\cdot\left(\rho_{0}{\bm{\nabla}} K_i\right) =  0$.

In one space dimension, we can in fact deduce the density $\varrho^{ss}$ exactly. Indeed the formula 
(\ref{varrhomatrix}) simplifies then into one term $\varrho^{ss} =  n - \frac{1}{\lambda} \int_{0}^\lambda  \rho_{0}(x)( \partial_x K_x)^2   \, d x  $ where $K_x(x) $ is a periodic function in the interval $[0,\lambda ]$ solution
of $\partial_x \rho_{0} +  \partial_x \left(\rho_{0} \partial_x K_x\right) =  0$. Thus $\partial_x K_x(x) = -1+ \frac{c}{\rho_{0}(x)}$ where $c$ is an integration constant. The periodic boundary condition
 $K_x(0) = K_x(\lambda) $ gives $c  = \frac{1}{\frac{1}{\lambda} \int_{0}^\lambda \frac{1}{ \rho_{0}(x) } dx}$.
  Finally, we find that in one dimension, the superfluid density writes:
 $$ \frac{1}{\varrho_{ss}} = \frac{1}{ c} =\frac{1}{\lambda} \int_{0}^\lambda \frac{1}{\rho_{0}(x) } dx.$$
 
Thus, the theory of homogenization provides us an exact result for the special case of one space dimension, and the effective density (scalar in 1D) is then a kind of ``harmonic" average of the density  \cite{lions}.
 
From this formula, the non classical rotational inertia fraction (NCRIF) $\varrho^{ss}/\rho$  corresponds exactly to the upperbound quotient $Q_0$
proposed by Leggett \cite{leggett}, who also established the equivalence for 1D systems more recently \cite{leggett2}.  Therefore, the NCRIF at low speed ($NCRIF_0$) reads:
\begin{equation}
\varrho^{ss}/\rho = Q_0\equiv \frac{1}{ \left( \frac{1}{\lambda} \int_0^\lambda  \rho_0(x) dx  \right)   \left( \frac{1}{\lambda} \int_0^\lambda  \frac{1}{ \rho_0(x) }  dx \right) }\label{rhoss}.
\end{equation}

\noindent {\bf \it Remarks.}

1. The Schwartz inegality\footnote{This inequality reads: $\left( \int_0^\lambda f(x)^2 dx \right)   \left( \int_0^\lambda g(x)^2 dx \right)  \geq \left( \int_0^\lambda f(x) g(x) dx \right)  ^2.$} and $\rho_0(x) \geq 0$ gives $0\leq Q_0 \leq 1.$

2. For finite energy, if  the ground state vanishes at some point, the non-classical rotational inertia does as well.
 Indeed, if at some point $x_*$ we have $\rho_0(x) \sim |x-x_*|^\alpha$ with $\alpha>0$, then
%\begin{eqnarray}\int_0^\lambda  \frac{1}{ \rho_0(x) }  dx & \approx & {\rm finite \ term} + \int_{x_* -\epsilon}^{x_*+\epsilon}   |x-x_*|^{-\alpha}dx \nonumber\\ &\approx &  {\rm finite \ term} + \frac{2}{1-\alpha} \epsilon^{1-\alpha},\label{rhoss=0}\end{eqnarray}
$$\int_0^\lambda  \frac{1}{ \rho_0(x) }  dx  \approx  {\rm finite \ term} + \int_{x_* -\epsilon}^{x_*+\epsilon}   |x-x_*|^{-\alpha}dx $$
and
$$Q_0 \approx \frac{1}{ {\rm finite \ term} + \frac{2}{1-\alpha} \epsilon^{1-\alpha} }.$$
Therefore if $0<\alpha \leq 1$, $Q_0$ remains finite with $\epsilon \rightarrow 0$. However, as we have seen above, such a ground state would require an infinite amount of energy.

\section{Results}

\subsection{ NCRIF in the weakly nonlinear limit.}

The NCRIF in the limit of weak modulation may be computed directly from the trial function (\ref{trial1}): 
\begin{equation}
Q_0 = \frac{(1-4|A|^2)^{3/2}}{(1+2 |A|^2)},
\label{rhossweak}
\end{equation}
where $|A|^2$ is evaluated at $k=k_c$. As $ |A| \rightarrow 1/2$, the quotient $Q_0$ vanishes, because the wavefunction (\ref{trial1}) vanishes at some point.

\subsection{ NCRIF in the  limit $\Lambda\rightarrow\infty$.}

For large $\Lambda$, since the ground state $\rho_0(x)$ decays exponentially, the contribution to
the NCRIF (\ref{rhoss}) mainly comes from the large contribution of $1/\rho_0(x)$ in 
$x\in [\delta,\lambda/2]$. That is, after using the WKB approximation:
\begin{equation}
Q_0 \approx \frac{5}{4}  K_*^2    e^{- \frac{\pi}{5}  \sqrt{\frac{ \Lambda \lambda }{3 \delta^3}} (\lambda/2-\delta) ^{5/2}   }. \label{Q0As}\end{equation}
\begin{figure}[hc]
\centering
({\it a})\includegraphics[width=8cm]{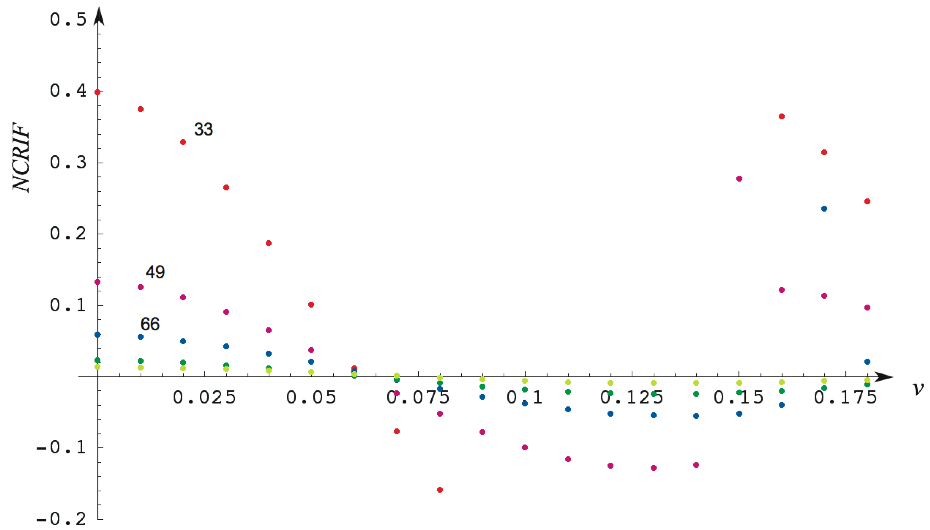} 
\hfill 
({\it b})\includegraphics[width=8cm]{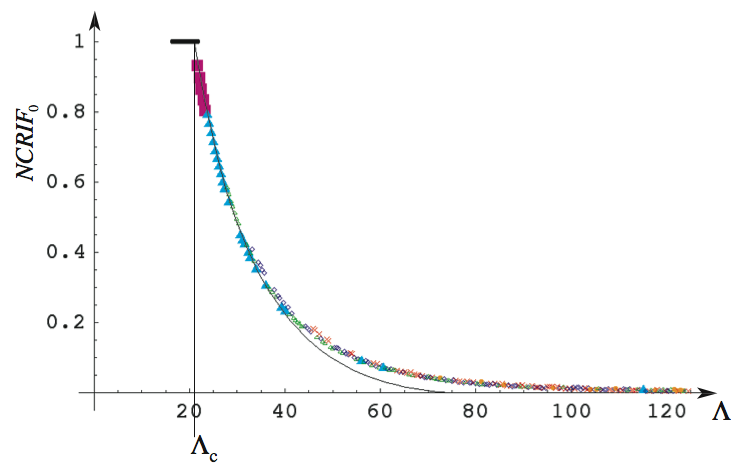}   
\caption{  \label{ncrif1da} (a)  $NCRIF$ as a function of $v$ for different values of $\Lambda$.  (b) $NCRIF_0$ as a function of $\Lambda$, the line is the curve from the weakly non linear analysis, see formula (\ref{rhossweak}) which gives a good approximation up to $\Lambda\approx 40$.}
\end{figure}

We will now be using numerical simulations to deduce the NCRI and compare it with theories by two different methods.
First, the ground state is determined. Then, one can compute directly the NCRI by imposing a
rotation to this ground state. On the other hand, the value of $Q_0$ can be calculated from the ground state solution $\rho_0(x)$.

Numerical results are obtained by minimizing the energy (\ref{densenergy}) under the number of particles condition (\ref{density}). We use therefore the Ginzburg-Landau version of the dynamics
which can be interprated as the integration of the GP equation for imaginary time $t=-i \tau$:
\begin{equation}
\frac{\partial\psi}{\partial \tau}  =  \mu \psi  + \frac{1}{2} \frac{ \partial^2 \psi}{\partial x^2}  - \frac{\Lambda}{2}   \psi(x,t)\int_{x-1}^{x+1} |\psi(y)|^2 {\mathrm{d}}y ,  
\label{minimization}
\end{equation}
$\mu$ is the Lagrangian multiplier introduced to satisfy the number of particles condition. 
%We start with a uniform solution perturbed either by a white noise or by a sinusoidal modulation in
%order to select $\lambda$.

Imposing a rotational frequency $\omega$ in a 1D annular system amounts to consider a drift
of the system at constant velocity $v=\omega L$ with periodic boundary conditions. The ground 
state of such a system is obtained by minimizing: ${\mathcal F} = {\mathcal E} + v{\mathcal P}+ \mu(N-n_0)$,
where ${\mathcal P}=-\frac{i}{2L} \int_{0}^{L}( \psi^*(x) \partial_x \psi(x) - \psi(x) \partial_x \psi^*(x) ) dx. $
Consequently, a direct computation of the NCRIF can be performed numerically:
$$NCRIF(v) = 1 - \frac{|{\mathcal P}'(v)|}{\int_0^L|\psi(x)|^2 dx}.$$
Fig. \ref{ncrif1da}(a) shows the function $NCRIF(v)$ for different $\Lambda$ obtained by numerical minimization of ${\mathcal F}$. As expected, the NCRIF decreases as
$v$ increases. For large value of the parameter $v$, the NCRIF first become negatives and then show
large fluctuations, indicating that complex structures are present, such as $2\pi$ phase jumps for instance
(similar to vortices in higher dimensions). Moreover, numerical instabilities are also enhanced by the
rotation so that only moderate $\Lambda$ (up to $150$) values could be achieved with full confidence.

The low speed limit: $$NCRIF_0 = \lim_{v\rightarrow 0} NCRIF(v)$$
is then shown on Fig. \ref{ncrif1da}(b) as function of $\Lambda$ and compared with the analytical
quotient (\ref{rhossweak}) of the weak amplitude modulations, showing an excellent agreement.

\begin{figure}[hc]
\centering
({\it a})\includegraphics[width=8cm]{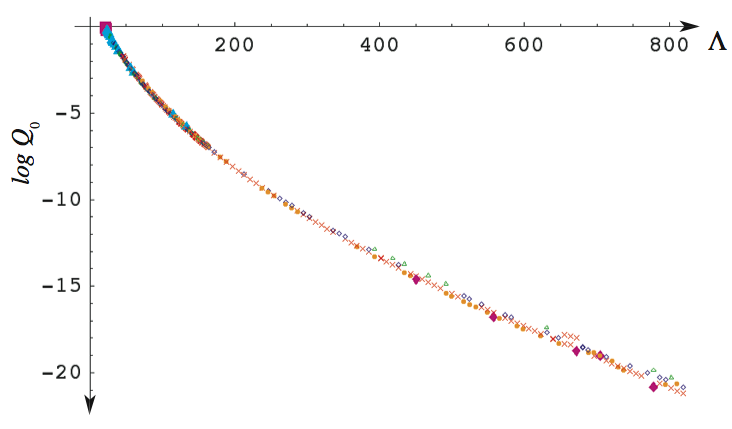} 
\hfill 
({\it b})\includegraphics[width=8cm]{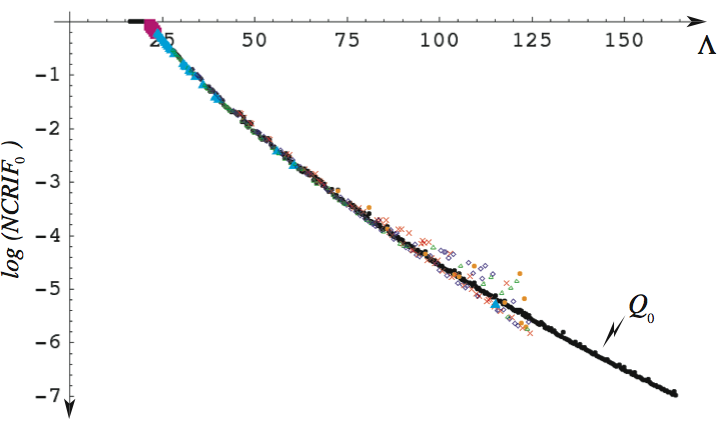}   
\caption{ \label{ncrifQ} (a) The quotient $Q_0$ as a function of $\Lambda$, using a direct numerical integration of ground states $\rho_0(x)$ obtained for Fig. \ref{deltavsLambda}. The lines are the functions obtained from the theory (\ref{Q0As}) for three different wavelength (similar notations as Fig. \ref{deltavsLambda}) with $K_* =0.01$ as a fixed parameter.  Notice the exponential behavior in qualitative agreement with (\ref{Q0As}) . (b) Comparison between the numerical calculations $NCRIF_0$ and the Leggett's quotient $Q_0$ represented with black dots.}
\end{figure}

On the other hand, as explained above, the $NCRIF_0$ can be calculated directly from the numerical
solution $\rho_0(x)$, by computing the Leggett quotient $Q_0$ (\ref{rhoss}).
Since the ground state solution is 
numerically more stable to obtain than the minimization of the rotating system, we are able to compute
 a satisfactory good estimates for $Q_0$ up to $\Lambda $ of the order of $800$, as
 shown on Fig. \ref{ncrifQ}(a).  Remarkably $Q_0$ does not depend on the wavelength of the periodic structure  $\lambda$, since all the numerical data for different $\lambda$ gather 
 on a single curve. This is a consequence that the main contribution to the quotient $Q_0$ comes from the wide region with small values of $\rho_0(x)$.  On the other hand, only poor agreement is found with the asymptotic behavior (\ref{Q0As}).

In Fig. \ref{ncrifQ}(b) we compare this quotient $Q_0$ with the $NCRIF_0$ obtained by direct numerical simulation of the rotating system for the accessible moderate $\Lambda$ values. It shows a particularly good numerical agreement between the two methods, as expected by the theory.

In conclusion, we have exhibited NCRI in a 1D model of supersolid in the context of annular geometry, using both direct numerical simulation and analytical estimates. In particular, we have shown that the 
so-called Leggett quotient was there in full agreement with NCRI.
C.J. acknowledges the financial support of the DGA for this research and this research was supported in part by the 
National Science Foundation under Grant No. PHY05-51164. S.R. acknowledges Anillo de Investigaci\'on Act. 15 (Chile).

\end{document}